\documentstyle[epsfig]{elsart}

\newcommand{\be}{\begin{equation}}
\newcommand{\ee}{\end{equation}}
\newcommand{\ba}{\begin{eqnarray}}
\newcommand{\ea}{\end{eqnarray}}
\begin{document}
\begin{frontmatter}
\title{Finite-Time Singularity Signature of Hyperinflation}
\author[iggp,ess,nice]{D. Sornette}, 
\author[sony]{H. Takayasu},
\author[iggp]{W.-X. Zhou}
\address[iggp]{Institute of Geophysics and Planetary Physics\\ University of California, Los Angeles, CA 90095}
\address[ess]{Department of Earth and Space Sciences\\ University of California, Los Angeles, CA 90095}
\address[nice]{Laboratoire de Physique de la Mati\`ere Condens\'ee,
CNRS UMR 6622 and Universit\'e de Nice-Sophia Antipolis, 06108 Nice Cedex 2, France}
\address[sony]{Sony Computer Science Laboratories,
3-14-13 Higashigotanda\\ Shinagawa-ku, Tokyo 141-0022, Japan\\
{\it E-mail addresses:}\/ sornette@moho.ess.ucla.edu (D.
Sornette), takayasu@csl.sony.co.jp (H. Takayasu),
wxzhou@moho.ess.ucla.edu (W.-X. Zhou)}
\begin{abstract}
We present a novel analysis extending the recent work of Mizuno et
al. \cite{MTT02} on the hyperinflations of Germany
(1920/1/1-1923/11/1), Hungary (1945/4/30-1946/7/15), Brazil
(1969-1994), Israel (1969-1985), Nicaragua (1969-1991), Peru
(1969-1990) and Bolivia (1969-1985). On the basis of a
generalization of Cagan's model of inflation based on the
mechanism of ``inflationary expectation'' or positive feedbacks
between realized growth rate and people's expected growth rate, we
find that hyperinflations can be characterized by a power law
singularity culminating at a critical time $t_c$. Mizuno et al.
\cite{MTT02} 's double-exponential function can be seen as a
discrete time-step approximation of our more general nonlinear ODE
formulation of the price dynamics which exhibits a finite-time
singular behavior. This extension of Cagan's model, which makes
natural the appearance of a critical time $t_c$, has the advantage
of providing a well-defined end of the clearly unsustainable
hyperinflation regime. We find an excellent and reliable agreement
between theory and data for Germany, Hungary, Peru and Bolivia.
For Brazil, Israel and Nicaragua, the super-exponential growth
seems to be already contaminated significantly by the existence of
a cross-over to a stationary regime.
\end{abstract}
\end{frontmatter}


\section{Introduction}

Inflation is the economic situation in which prices apparently
move monotonically upward and the value of money decreases. To
classical economics, inflation is the undue increase in the supply
of credit above the level that is supported by current savings.
High inflation is always associated with high rates of money
supply growth while the relationship is weak for countries with
low inflation \cite{degrauwe}. Thus, fighting high inflation
requires reducing the growth rate of the money supply.

Inflation is one of the few big issues in macroeconomics, together
with unemployment, monetary policy, fiscal policy, import-export
deficits, productivity, government spending and the business
cycle, and has been at the forefront of public battles over the
past half-century. A good economic policy should strive to achieve
a balance between often contradictory requirements: for instance,
many economists assume that unemployment tends toward a natural
rate below which it cannot go without creating inflation.
Samuelson and Solow had brought to the U.S. the empirical
evidence, first compiled by the British economist A.W. Phillips,
that there seems to be a tradeoff between inflation and
unemployment--that is, higher inflation meant lower unemployment.
There is thus a long tradition among economists to adopt monetary
policy as a way to keep the economy running on high-employment
overdrive. Allowing prices to rise seemed the only humane thing to
do. Friedman argued however that the unemployment/inflation
tradeoff was temporary, and he also pointed out that using fiscal
and monetary policy to avert recessions was a lot harder than it
looked. The difficulties stem from the fact that policies designed
to restrain inflation by lowering the level of aggregate demand
will tend to depress investment and harm capacity. Improved
industrial performance requires a climate conducive to investment
and research and development, which in turn depends on, inter
alia, high and stable levels of aggregate demand. Business and
inflation cycles often result from the combination of endogenous
interactions (that can lead to incoherence) and of the effects of
institutions to contain these tendencies in the economy. The
corresponding economic times series can exhibit smooth growth and
well-behaved cycles as possible transitory results of the economic
processes, but can also allow for intermittent conditions
conducive to the emergence of incoherence or turbulence.
Institutional factors attempt to act as circuit breakers on the
economy. Whenever institutionally determined values dominate
endogenously determined values, the path of the economy is broken
and an interactive process, which starts with new initial
conditions, generates future values. Specifically, whenever the
economy threatens to behave incoherently, these stabilizers,
whether built-in or activated by government authority, prevent the
economy from continuing on the prior determined path, with the
corresponding added complication and possible elements of
destabilization. These are important elements in the path
evolution of inflation.

In standard economic theory, inflation is associated with money
supply growth. At equilibrium, money determines price level and
implies equilibrium in markets for other assets. At equilibrium,
money demand depends primarily on income and interest rates. But
there are several factors keeping money demand unstable, such as
financial innovations as well expectations. Indeed, one of the
major causes of the complexity in stabilizing inflation together
with other macroeconomic variables is that expectations of
producers, consumers and investors may play a key role in the
dynamics. Indeed, investment allocations or inflation expectations
are influenced by ex-ante values of the risk premia and ex-post
returns are rough approximations of these. Thus, ``inflationary
expectation'' occurs when people begin to raise prices not because
of actual changes in supply or demand or cost or the size of the
money supply, but out of fear that some such changes might happen.
In the 1990s, when Alan Greenspan, the chairman of the US federal
reserve, said that the U.S. was still suffering from the
inflationary expectations caused by the monetary excess of the
1970s, he was directly addressing the potential for inflation
caused by ``inflationary expectations.'' When European central
banks added liquidity to the gold market in an attempt to prevent
an increase in the price of gold from creating concerns about a
decrease in the value of the dollar, they were addressing the
psychological component of price stability involved in
``inflationary expectations.'' Mathematically, this dynamics
translates into sets of coupled nonlinear equations expressing
both the competition and delays between expectations and
realizations and the presence of positive and negative feedback
loops. The complexity of the resulting dynamics stems from the
complex nonlinear negative and positive feedback processes
intertwining the different component of policies.

There are several causes of inflation. A prominent origin is wars,
which cause the type of inflation that results from a rapid
expansion of money and credit. For instance, in World War I, the
American people were characteristically unwilling to finance the
total war effort out of increased taxes. This had been true in the
Civil War and was also so in World War II and the Vietnam War.
Much of the expenditures in World War I, were financed out of the
inflationary increases in the money supply. If money supply growth
and real income are constant, then expected inflation rate equals
current inflation rate (assuming no change in elasticities). This
is more or less the standard situation most of the time, as
nominal interest rates and inflation often move together. In
contrast, if people expect an increase in money growth, this then
would lead to expect higher inflation. And expectation of higher
inflation raises inflation rate even if money growth does not
actually increase.

If inflation is perfectly anticipated, it entails no cost for
creditors and debtors as nominal interest rates incorporate
expected inflation and nominal wages adjust to offset price
increases. But inflation devalues the currency and imposes ``shoe
leather costs'', that are costs of efforts to minimize cash
holding (for instance the time and effort in making lots of trips
to ATM machines). Prices will be changed more frequently and this
imposes ``menu costs,'' which are the  costs of changing prices.

If inflation is unanticipated, it induces transfers of wealth from
holders of nominal assets to holders of real assets \cite{temple}.
Suppose for instance that your savings account pays 8\% per year
and that you expected 4\% inflation but the realized inflation is
7\%.  You obtain a real interest rate is 1\% instead of the 4\%
that you expected. You are worse off but the bank is better off.
Unanticipated inflation increases risk of gaining or losing wealth
and requires more resources for forecasting inflation.
Unanticipated inflation causes confusion about the relative price
movements as it could affect some prices sooner than others. What
if the price of oil increases relative to natural gas?  Is that a
change in relative prices, or a result of inflation? If the former
holds, consumers should switch from oil to natural gas for
heating. If the latter holds, and they switch, then resources are
misallocated. More generally, informal accounts of inflation's
effects are common, but there are few models which get to grips
with the central effects. Partly as a result of this, and partly
as a result of many econometric problems, much of the empirical
evidence remains unconvincing (see \cite{temple} for an assessment
of the various contributions). For all these reasons, a main
target of central banks of developed countries in the last decade
of the twentieth century has been a low inflation \cite{Froyen}.

As we have seen, inflation is first-of-all an indirect tax
leveraged by governments through their (partial) control of the
money supply to help them finance wars or other expenditures. The
problem is that inflation is not easily controlled due to the dual
effect of financial innovations and expectations. Once people
start to expect an inflation regime, their expectations may lead
to strong positive feedbacks that make inflation run away. There
are several remarkable historical examples of such runaways,
called ``hyperinflation,'' such as those that occurred in Germany
(1922-1923), Hungary (1945-1946), Latin America in the 1980s and
Russia in the recent years. Such hyperinflation phases are very
costly to society, as there are enormous ``shoe-leather'' costs,
the workers have to be paid more frequently (even daily) and there
are rushes to spend the currency before prices rise further.
Hyperinflation reduce real value of taxes collected, which are
often set in nominal terms and by the time they are paid, real
value has fallen. Hyperinflation leads to large disruptive effects
on price and on wage changes and prevents distinguishing relative
from aggregate price movements. Wealth allocation becomes very
inefficient. Detecting hyperinflation in an early stage might
contribute to avoid such tragedy.

In a recent work, Mizuno et al. \cite{MTT02} have analyzed the
hyperinflations of Germany (1920/1/1-1923/11/1), Hungary
(1945/4/30-1946/7/15), Brazil (1969-1994), Israel (1969-1985),
Nicaragua (1969-1991), Peru (1969-1990) and Bolivia (1969-1985),
and showed that the price indices or currency exchange rates of
these countries grew super-exponentially according to a
double-exponential function $e^{b_1 e^{b_2 t}}$ of time (with
$b_1, b_2 >0$). This super-exponential growth was argued to result
from a nonlinear positive feedback process in which the past
market price growth influences the people's expected future price,
which itself impacts the ex-post realized market price. This
autocatalytic process is fundamentally based on the mechanism of
``inflationary expectation'' alluded to above and is similar to
the positive feedbacks occurring during speculative bubbles due to
imitative and herd behaviors \cite{bookcrash}.

Clearly, a super-exponential growing inflation is unsustainable.
While providing a useful mathematical description of
hyperinflation, the double-exponential model of Mizuno et al does
not provide a rigorous determination of the end of the
hyperinflation regime \cite{MTT02}. Here, we re-examine the theory
and empirical evidence developed in \cite{MTT02} and show that the
double-exponential law is nothing but a discrete-time
approximation of a general power law growth endowed with a
finite-time singularity at some critical time $t_c$. The
finite-time singularity allows us to define unambiguously the
theoretical end of the hyperinflation regime as being $t_c$ by
definition. This theory provides the first practical approach for
detecting hyperinflation and predicting its future path until its
end. In practice, the end of an hyperinflation regime is expected
to occur somewhat earlier than at the asymptotic critical time
$t_c$, because governments and central banks are forced to do
something before the infinity is reached in finite time. Such
actions are the equivalent of finite-size and boundary condition
effects in physical systems undergoing similar finite-time
singularities. Hyperinflation regimes are of special interest as
they emphasize in an almost pure way the impact of collective
behavior of people interacting through their expectations.

\section{From double-exponential growth to finite-time singularity}

In order to establish the correspondence between the
double-exponential and the power law formulas, let us first
summarize the approach followed by Mizuno et al. \cite{MTT02} who
extend Cagan's theory of inflation \cite{Cagan} in terms of a set
of evolution equations linking the market price $p(t)$ with the
people's averaged expectation price $p^*(t)$. These two prices are
thought to evolve due to a positive feedback mechanism: an upward
change of market price $p(t)$ in a unit time $\Delta t$ induces a
rise in the people's expectation price, and such an anticipation
pulls up the market price. Cagan's assumption that the growth rate
of $p^*(t)$ is proportional to the past realized growth rate of
the market price $p(t)$ is expressed by the following equation \ba
{p(t+\Delta t) \over p(t)} &=& {p^*(t)  \over p(t)}~,  \label{mgnsl} \\
{p^*(t+\Delta t) \over p^*(t)} &=& {p(t)  \over p(t-\Delta t)}~.  \label{mgnslfgd}
\ea
Introducing the growth rates defined by
\ba
r(t) &\equiv & {p(t+\Delta t) \over p(t)}~, \\
r^*(t) &\equiv & {p^*(t+\Delta t) \over p^*(t)}~,
\ea
expressions (\ref{mgnsl}) and (\ref{mgnslfgd}) are equivalent to
\ba
r(t) &=& r^*(t-\Delta t)~,   \label{nbjgls}\\
r^*(t) &=& r(t-\Delta t)~,   \label{njjkjd}
\ea
whose solution is $r(t+\Delta t) = r(t-\Delta t)$ and expresses
the spontaneous formation of a constant finite growth rate characterizing
a steady state exponential inflation.

In order to explain hyperinflation regimes, Mizuno et al.
\cite{MTT02} generalizes Cagan's assumption that the growth rate
of $p^*(t)$ is proportional to the past realized growth rate of
the market price $p(t)$ and introduce a nonlinear dependence in
equation (\ref{mgnslfgd}) which is transformed into \be
{p^*(t+\Delta t) \over p^*(t)} = \left({p(t) \over p(t-\Delta
t)}\right)^{b}~,  \label{mgnaaslfgd} \ee which can be written \be
r^*(t) = b ~r(t-\Delta t)~. \label{mgalz} \ee Cagan's original
model is recovered for the special case $b=1$. The system
(\ref{mgnsl},\ref{mgnaaslfgd}) or equivalently
(\ref{nbjgls},\ref{mgalz}) studied by Mizuno et al. \cite{MTT02}
is obtained from a coarse-graining (or Monte-Carlo renormalization
group) procedure of a more general system of equations developed
by Mizuno et al. \cite{MTT02}.

Expression (\ref{mgnsl}) describes the dynamical tendency for the
market price $p(t)$ to converge towards the expected price
$p^*(t)$. The equation (\ref{mgnaaslfgd}) involving the nonlinear
function $(p(t)/p(t-\Delta t))^{b}$ with $b>1$ captures the
concept that people re-adjust their expectation in a significant
way only if the realized market price change in the previous unit
time interval was significant. An exponent $b$ larger than $1$
captures the fact that the adjustment of the expected price $p^*$
is weak for small changes of the realized market prices and
becomes very strong for large deviations. This embodies the fact
that people have only a rough sense of what to expect for the
future and can only form rather fuzzy expectations. The expected
price $p^*(t)$ is thus estimated with a rather poor credibility.
As a consequence, the agents have no incentive to move much their
expectation if the past realized prices have changed little
because such small change might be within the fuzziness zone of
determination of their expectations. Only when the price change is
large, will the agents act and modify their expectation of a large
future increase of the prices, thus making it happen.  This effect
is embodied in the nonlinear function $(p(t)/p(t-\Delta t))^{b}$
with $b>1$, leading to a kind of threshold effect. The larger is
the exponent $b$ above $1$, the closer is this effect to a
threshold below which the people's expectation do not change much
and above which the revision of their expectation is dramatically
revised upward. We believe that such nonlinear response functions
embody much more realistically real people's behavior than do
linear models used in standard economic models. Exactly the same
mechanism has been invoked in a dynamical model of market prices
resulting from the interplay between fundamental and technical
investments in Refs.\cite{IS02,SI02}.

The system (\ref{nbjgls},\ref{mgalz}) gives
$r(t+\Delta t) = b ~r(t-\Delta t)$, whose solution is $r(t) \propto e^{b_2 t}$
leading to the announced
double exponential form for the market price \cite{MTT02}
\be
p(t) \approx e^{b_1 e^{b_2 t}}~,
\label{mklws}
\ee
where $b_1$ and $b_2$ are two positive constants.

Here, we propose an different version of the nonlinear feedback
process. We keep expression (\ref{mgnsl}) or equivalently equation
(\ref{nbjgls}) and replace equation (\ref{mgnaaslfgd}) or
equivalently expression (\ref{mgalz}) by \be r^*(t) = r(t-\Delta
t) + a [r(t-\Delta t)]^m~,~~~{\rm with}~~m > 1~. \label{msagalz}
\ee Note that our new formulation (\ref{msagalz}) retrieves
Cagan's formulation (\ref{njjkjd}) for $a=0$. It is also close to
Mizuno et al.'s form \cite{MTT02}, which is recovered for $m=1$.
We believe that this formulation (\ref{msagalz}) better captures
the intrinsically nonlinear process of the formation of
expectations. Indeed, if $r(t-\Delta t)$ is small (explicitly, if
it is smaller than $1/a^{1/m}$), the second nonlinear term $a
[r(t-\Delta t)]^m$ in the right-hand-side of (\ref{msagalz}) is
negligible compared with the first Cagan's term $r(t-\Delta t)$
and one recovers the exponentially growing inflation regime of
normal times. However, when the realized growth rate becomes
significant, people's expectations start to amplify these realized
growth rates, leading to a super-exponential growth.
Geometrically, the difference between our formulation
(\ref{msagalz}) and that of Mizuno et al.'s form \cite{MTT02}
consists in replacing a straight of slope $b$ larger than $1$ by a
upward convex function with unit slope at the origin and whose
local slope increases monotonously with the argument.

Putting equation (\ref{nbjgls})
together with expression (\ref{msagalz}) leads to
\be
r(t+\Delta t) =
r(t-\Delta t) + a [r(t-\Delta t)]^m~.
\label{mgmlzsz}
\ee
Keeping time discrete, the long-time solution of
(\ref{mgmlzsz}) is dominated by the second term $a [r(t-\Delta t)]^m$
for $m>1$ and is of the form
\be
r(t) \sim (r_0)^{m^{(t/2\Delta t)}}
\label{mgjlde}
\ee
 for some constant $r_0$, that is, it
takes the form of a double-exponential growth for the
growth rate and thus of a triple exponential growth for the market
price. Taking the continuous limit, expression (\ref{mgmlzsz})
becomes
\be
{d r \over dt} = a_1 [r(t)]^m~,
\label{mgmlsl}
\ee
where $a_1$ is a positive coefficient. Its solution exhibits a
finite-time singularity
\begin{equation}
r(t) = a_1 r(0) \left(\frac{t_c}{t_c-t}\right)^{1/(m-1)}~,
\label{eq:xfts}
\end{equation}
where the critical time $t_c = (m-1)/[r(0)]^{m-1}$ is
determined by the initial condition $r(0)$ and the exponent $m$.
The comparison between expressions (\ref{mgjlde}) and (\ref{eq:xfts})
reveals the general fact that a finite-time singularity (\ref{eq:xfts})
becomes a double-exponential (\ref{eq:xfts}) when the dynamics is
expressed in discrete time steps. Indeed,
in contrast with a continuous ordinary differential
equation (ODE) which may exhibit a finite-time singularity,
a discrete equation can not exhibit a genuine
exact finite time singularity. A true finite time
singularity is impossible as soon as time is discrete.
The reason is clear: the
finite time singularity comes from the fact that the doubling time
is divided by some factor larger than one after a time evolution
equal to the doubling time, so that the doubling
time shrinks to zero eventually (this is a simple way
to view the finite time singularity). When the doubling time
becomes comparable to the time step $\Delta t$, it cannot shrink
below it and there is a crossover from the finite
time singularity acceleration to an un-ending exponential of
exponential growth.
Thus, a power law singularity is essentially undistinguishable
from an exponential of an exponential of time, except when the
distance $t_c-t$ from the finite time singularity becomes
comparable with the time step $\Delta t$. This is the reason
why our present analysis is compatible with that reported in
Ref.~\cite{MTT02}. The main difference lies in the fact that the
continuous time solution contains an information on the end of the
growth phase, embodied in the existence of the critical time $t_c$.

The price is the exponential of the integral of $r(t)$
and also exhibits a finite-time singularity at the same critical
value $t_c$. The time dependence of the market price $p(t)$
exhibits the two following regimes.
\begin{itemize}
\item Finite-time singularity in the price itself:
\begin{equation}
\ln p(t) = A + B \left({t_c-t}\right)^{-\alpha},~{\rm with}~
\alpha = {2-m \over m-1} >0 ~~{\rm and}~B>0~, ~{\rm for}~1 < m < 2.
\label{eq:pure}
\end{equation}
This solution corresponds to a genuine divergence of $\ln p(t)$ in
finite time at the critical value $t_c$.
\item Finite-time singularity in the derivative or slope of the price:
\begin{equation}
\ln p(t) = A' - B' \left({t_c-t}\right)^{\alpha'},~{\rm with}~
\alpha' = {m-2 \over m-1} >0 ~{\rm and}~~B'>0~, ~{\rm for}~2 < m~.
\label{eq:pure22}
\end{equation}
As time approaches the critical value $t_c$, the price accelerates
with an infinite slope
(since $0< \alpha'<1$ for $m>2$) reached
at $t_c$, while remaining finite at the value $A'$.
\end{itemize}
$A$ and $A'$ are additive constants resulting from
the integration of the growth rate $r(t)$.
Such finite-time singularities are similar to those obtained
in the dynamics of the
world population, economic and financial indices \cite{JS01} and
can be seen as a special case of the Sornette-Ide model \cite{SI02,IS02}.

We will thus use equations (\ref{eq:pure}) and (\ref{eq:pure22})
to fit the hyperinflation of the historical price index of
Bolivia, Peru, Israel, Brazil, Nicaragua, Hungary and Germany
(ordered from the most recent to most ancient).

In addition, in order to test the robustness of these power laws, it may be useful
to recognize that the whole time span may not be fully captured by the
inflationary expectation mechanism embodied in expression (\ref{mgmlzsz})
and that the power law finite-time singularities may be preceded by
a non-singular regime. We postulate that the early time non-singular
regime may be described by replacing $t_c-t$ in equation (\ref{eq:pure})
by $\tanh[(t_c-t)/\tau]$, which describes a crossover from exponential growth
to a power law singularity. This amounts to replacing
expressions (\ref{eq:pure}) or (\ref{eq:pure22}) by
\begin{equation}
\ln[p(t)] = A + B \tanh[(t_c-t)/\tau]^{-\alpha}~, \label{eq:tanh}
\end{equation}
which contains a novel parameter $\tau$. Notice that, when $t_c-t
\ll \tau$, $\tanh[(t_c-t)/\tau] = (t_c-t)/\tau$ and one recovers
the pure power law (with $\tau$ digested inside the constant $B$).
Only for $t_c-t \geq \tau$ does the hyperbolic tangent provide a
cross-over to an exponential law. This form (\ref{eq:tanh}) has
been found to describe very well the cross-over from the
non-critical to the critical regime of rupture of heterogeneous
materials \cite{SA98}, allowing a significant improvement of the
reliability of failure predictions \cite{JS00}.

\section{Applications}

\subsection{Clear-cut cases of finite-time singularities}

\begin{figure}
\begin{center}
\epsfig{file=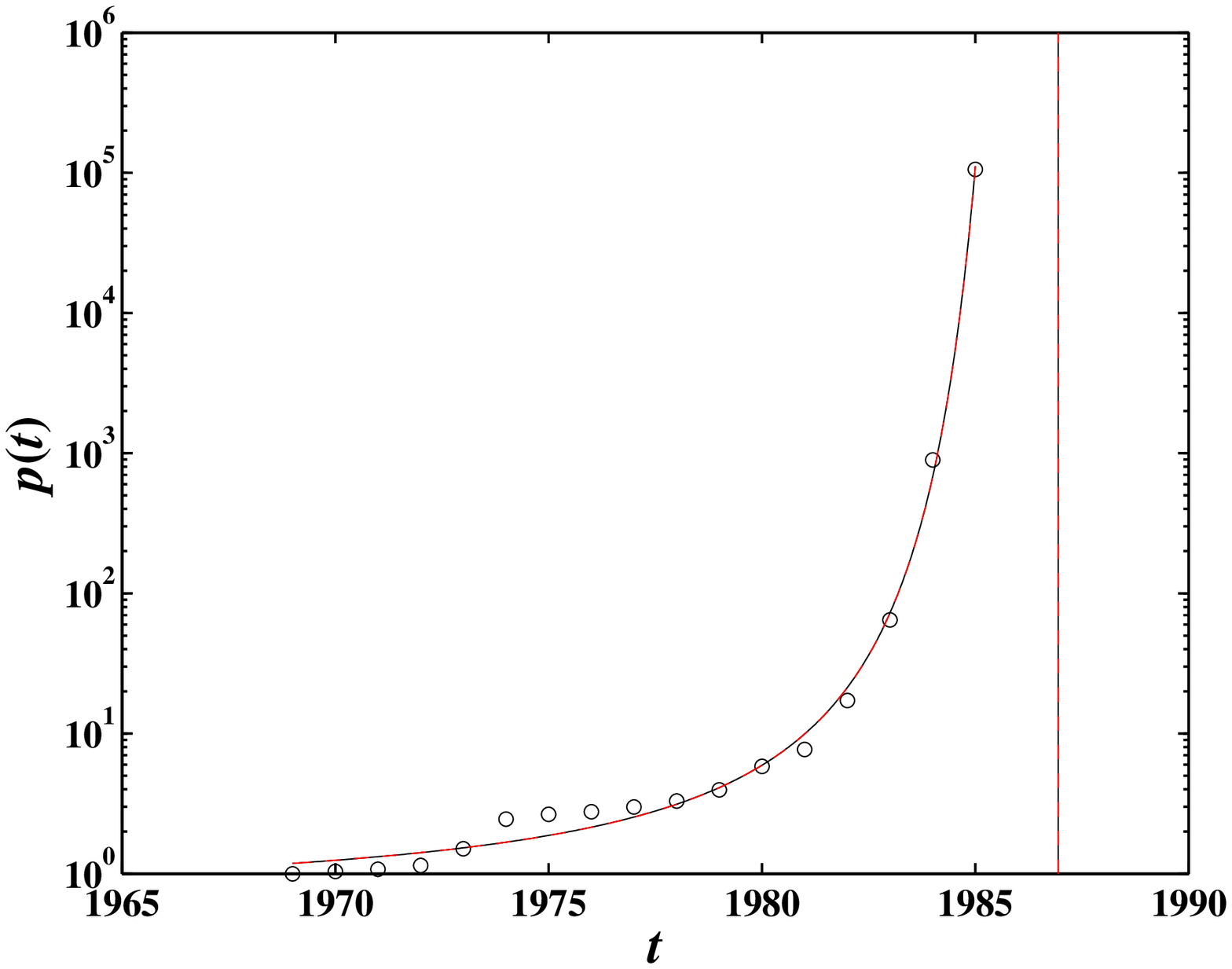,width=13cm}
\end{center}
\caption{Semi-logarithmic plot of the yearly price index of
Bolivia from 1969 to 1985 marked with open circles and its fits with
the finite time singular functions (\ref{eq:pure}) (solid line) and
(\ref{eq:tanh}) (dashed line). The vertical lines indicate
the corresponding predicted critical time $t_c$.} \label{Figbol}
\end{figure}

\begin{figure}
\begin{center}
\epsfig{file=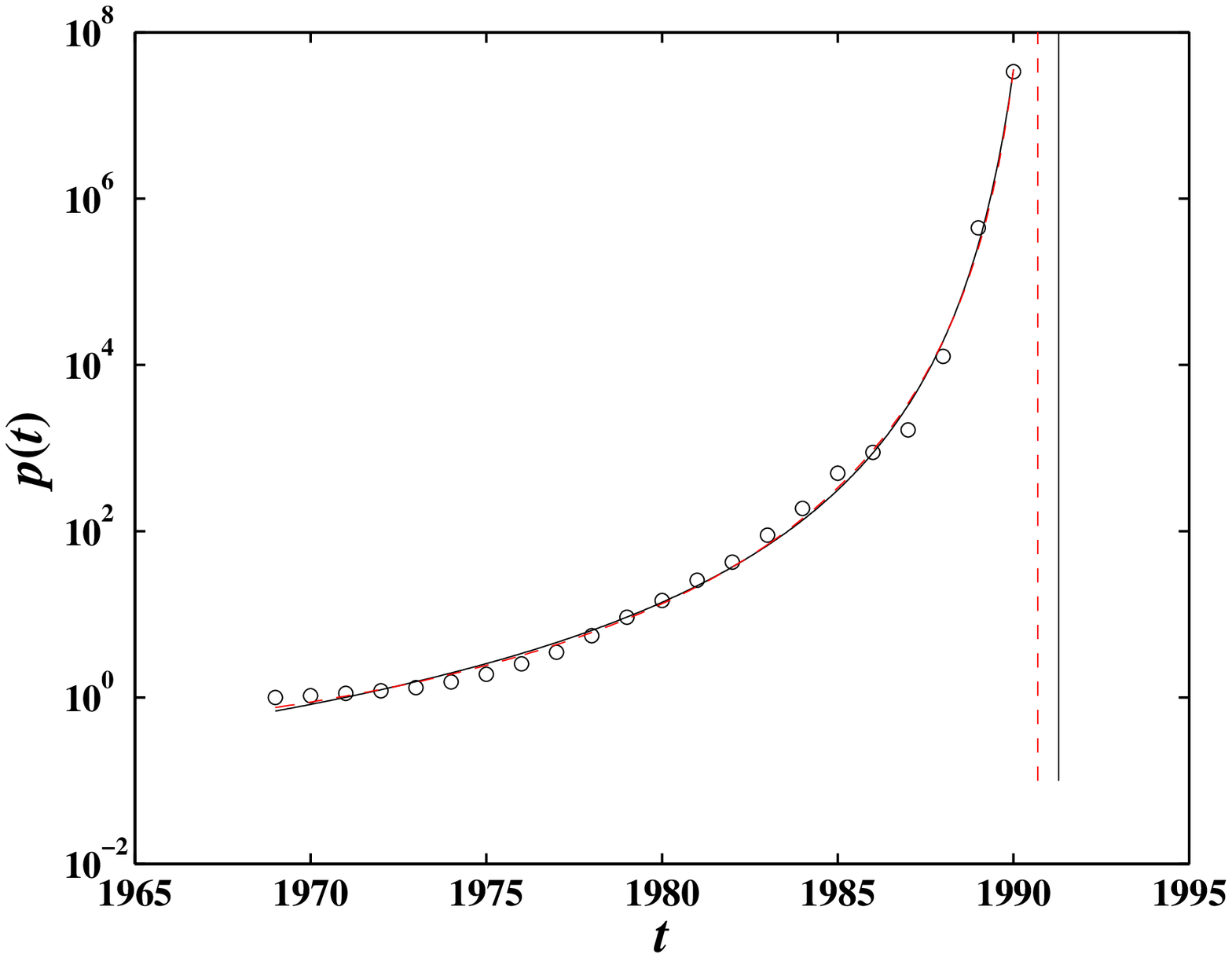,width=13cm}
\end{center}
\caption{Semi-logarithmic plot of the yearly price index of Peru
from 1969 to 1990 marked with open circles and its fits with the finite
time singular functions (\ref{eq:pure}) (solid line) and
(\ref{eq:tanh}) (dashed line). The vertical lines indicate
the corresponding predicted critical time $t_c$.} \label{Figperu}
\end{figure}

\begin{figure}
\begin{center}
\epsfig{file=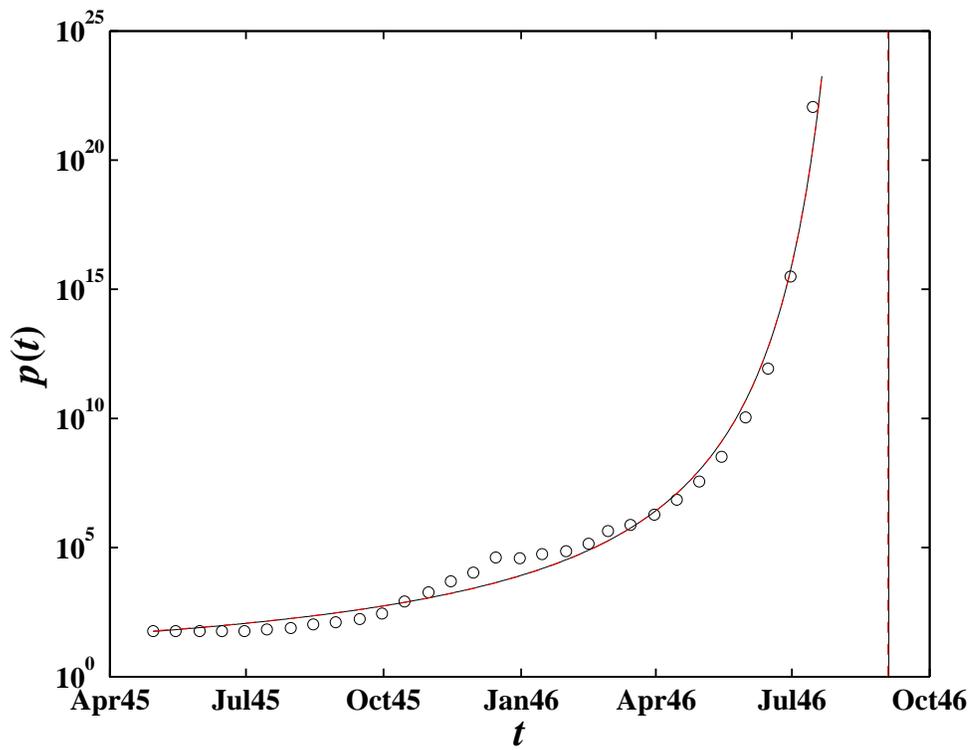,width=13cm}
\end{center}
\caption{Semi-logarithmic plot of the price index of Hungary from
April 30, 1945 to July 31, 1946 marked with open circles and its
fits with the finite time singular functions (\ref{eq:pure}) (solid
line) and (\ref{eq:tanh}) (dashed line). The vertical lines indicate
the corresponding predicted critical time $t_c$.} \label{Fighung}
\end{figure}

\begin{figure}
\begin{center}
\epsfig{file=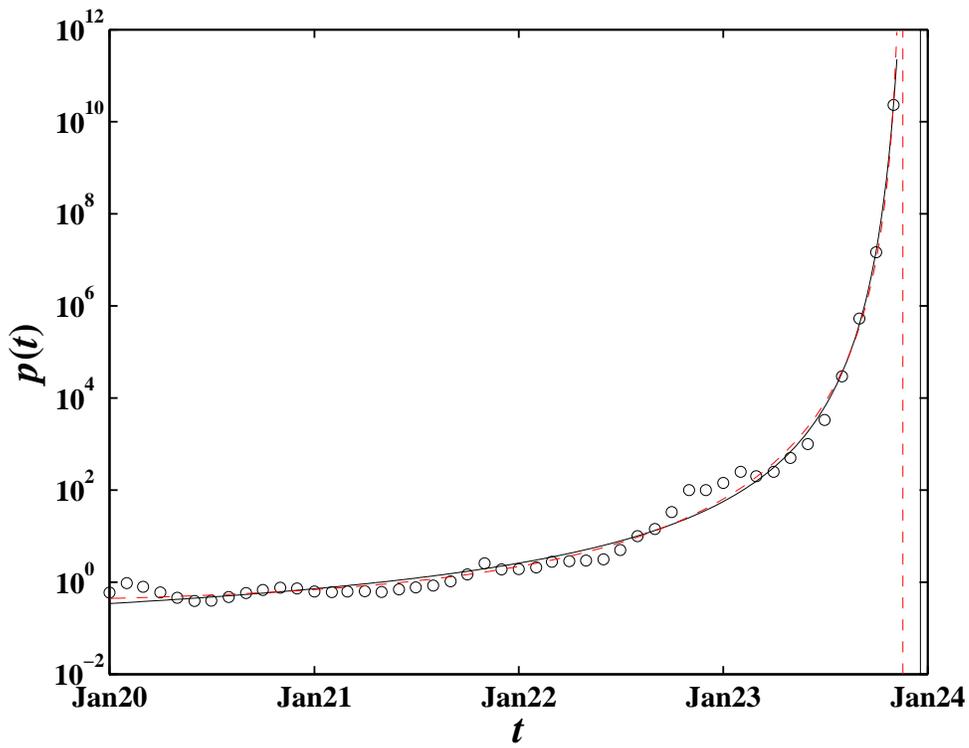,width=13cm}
\end{center}
\caption{Semi-logarithmic plot of the price index of Germany from
January 1920 to November 1923 marked with open circles and its
fits to the finite time singular functions (\ref{eq:pure}) (solid
line) and (\ref{eq:tanh}) (dashed line). The vertical lines indicate
the corresponding predicted critical time $t_c$.} \label{Figgerm}
\end{figure}

The hyperinflations of Bolivia, Peru, Hungary and Germany are well
fitted by expressions (\ref{eq:pure}) (continuous lines) and
(\ref{eq:tanh}) (dashed lines) as shown in Figs.~\ref{Figbol},
\ref{Figperu}, \ref{Fighung} and \ref{Figgerm}, respectively. The
parameters of the fits with formulas (\ref{eq:pure}) and
(\ref{eq:tanh}) to the hyperinflation price time series of the
four countries Bolivia, Peru, Hungary and Germany are given in
Table \ref{4function}. There are very small differences between
the fits obtained with expressions (\ref{eq:pure}) and
(\ref{eq:tanh}), suggesting that the considered time intervals are
fully in the inflationary expectation regime with strong positive
feedbacks. In particular, the exponents $\alpha$ are very robust
and the critical times $t_c$ are unchanged between the two
formulas for Bolivia and Hungary, while $t_c$ is moved by two
weeks for Germany and by four months for Peru. We also tested the
robustness of these results by restricting the fits with to the
two formulas (\ref{eq:pure}) and (\ref{eq:tanh}) to the last half
of each time series. We find that the exponents $\alpha$ and
critical times $t_c$ are essentially unchanged for the two most
dramatic hyperinflation of Hungary and Germany, while $t_c$ is
pushed forward in the future for Bolivia (by a few years) and Peru
(by a few months), without significant degradation of the quality
of the fits. This shows that only for Hungary and Germany can one
ascertain the critical time of the finite-time singularity with
good precision.

In the case of Hungary, the hyperinflation was eventually stopped
by the introduction of the present Hungary currency, the Forint,
in July 1946. Our prediction of the critical time $t_c$ at the
beginning of September 1946 suggests that an action on the part of
the government was unavoidable as the hyperinflation was close to
its climax.

\subsection{Evidence of a finite-time singularity regime in
$p(t)$ and not in $\ln p(t)$}

\begin{figure}
\begin{center}
\epsfig{file=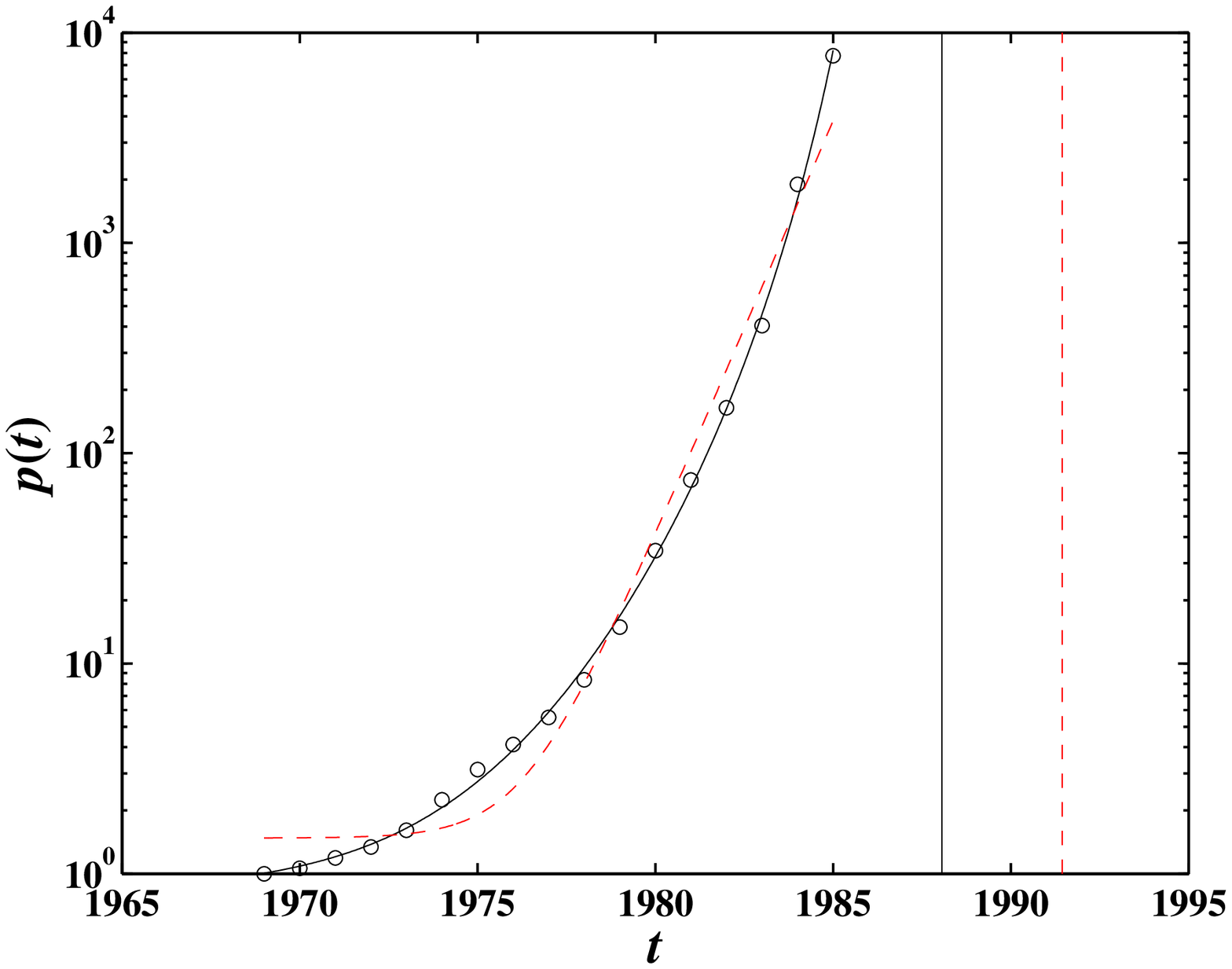,width=13cm}
\end{center}
\caption{Semi-logarithmic plot of the yearly price index of Israel
from 1969 to 1985 marked with open circles and its fits to the right-hand-side of
expressions (\ref{eq:pure}) (solid line) and of
(\ref{eq:tanh}) (dashed line).} \label{Figisr}
\end{figure}

\begin{figure}
\begin{center}
\epsfig{file=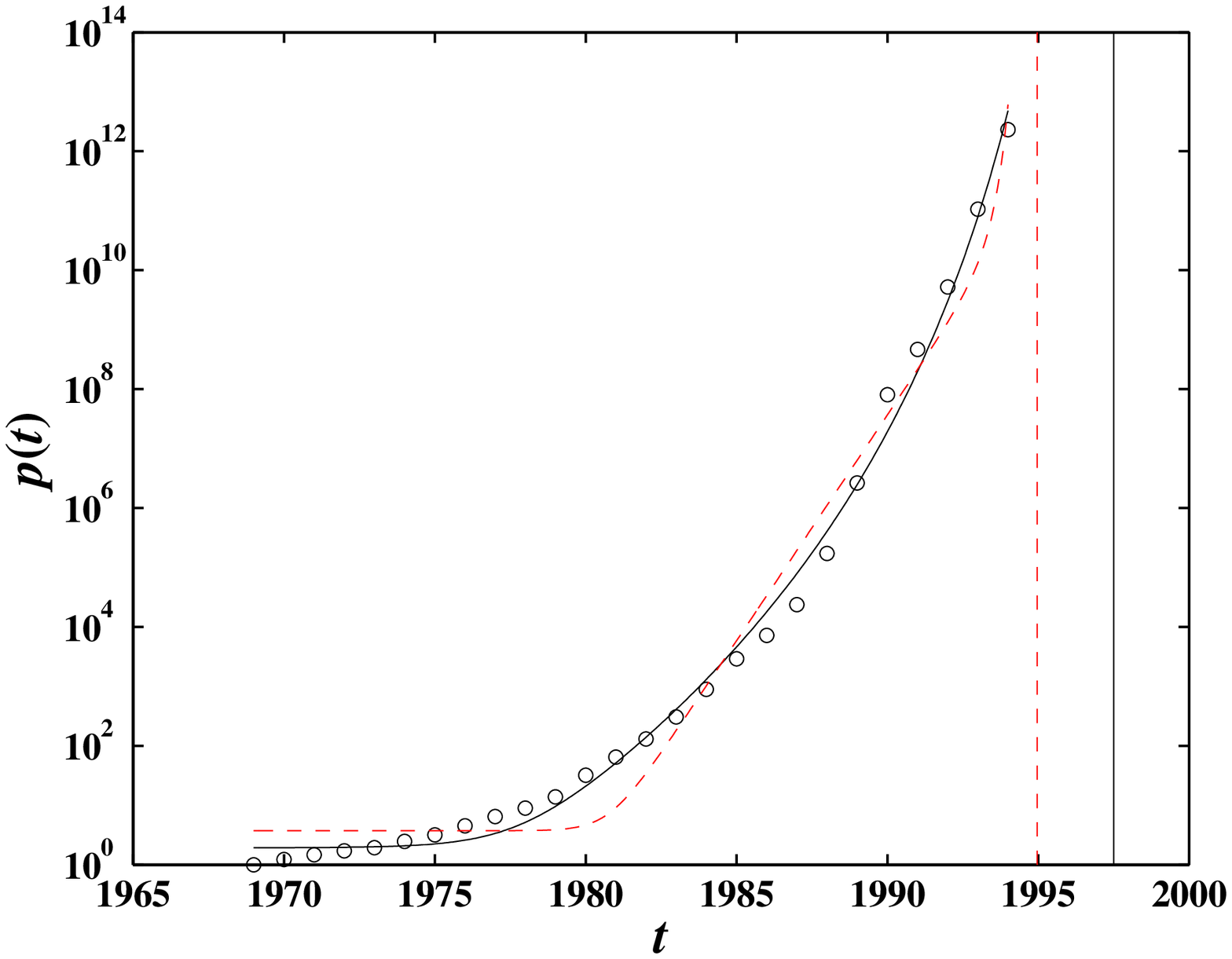,width=13cm}
\end{center}
\caption{Semi-logarithmic plot of the yearly price index of Brazil
from 1969 to 1994 marked with open circles and its fits to the right-hand-side
of expressions (\ref{eq:pure}) (solid line) and
(\ref{eq:tanh}) (dashed line).} \label{Figbra}
\end{figure}

\begin{figure}
\begin{center}
\epsfig{file=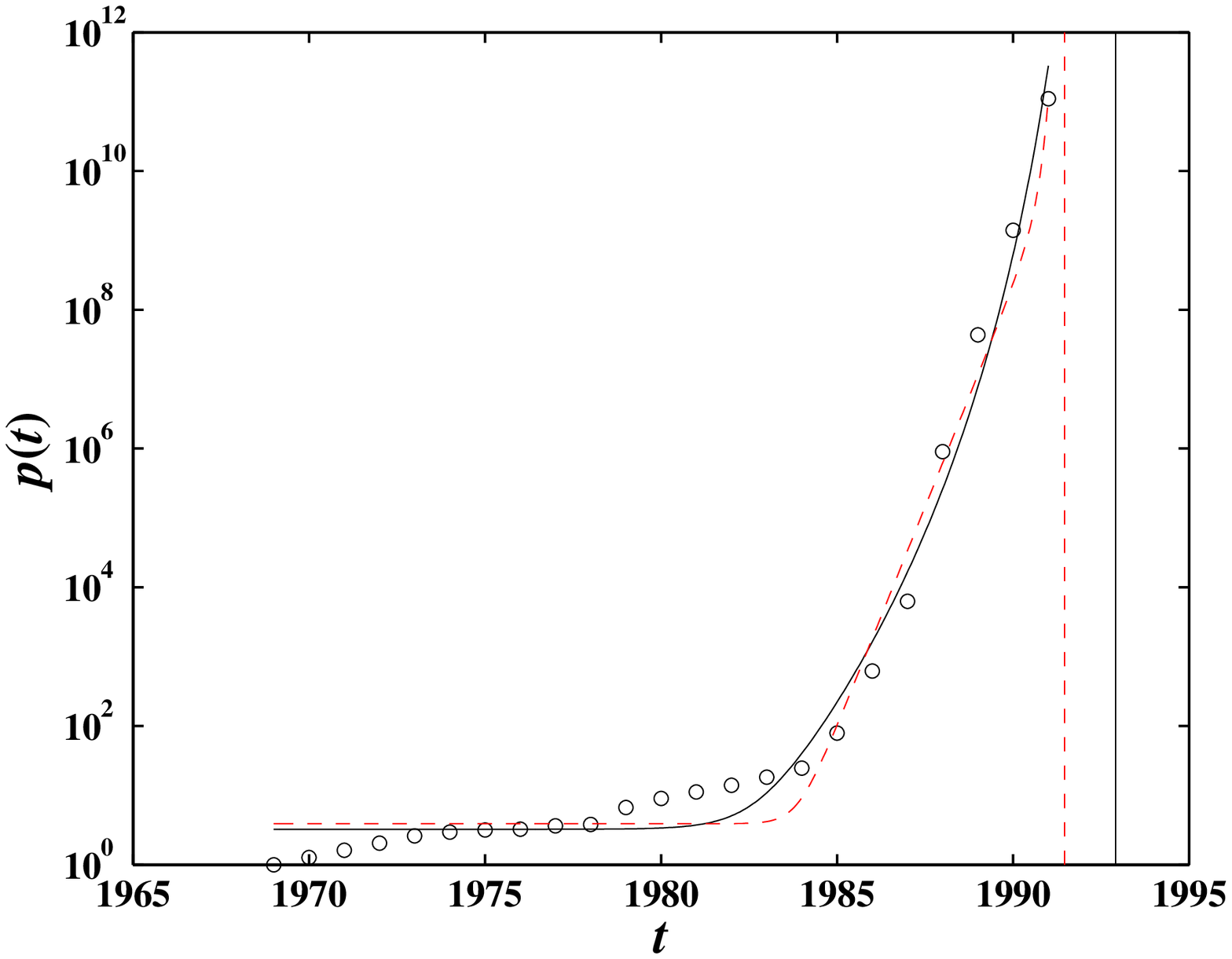,width=13cm}
\end{center}
\caption{Semi-logarithmic plot of the yearly price index of
Nicaragua from 1969 to 1991 marked with open circles and its fits
to the right-hand-side of expressions (\ref{eq:pure}) (solid line)
and (\ref{eq:tanh}) (dashed line).} \label{Fignica}
\end{figure}

The cases of Israel, Brazil and Nicaragua are not as clear-cut.
While the hyperinflation of these countries clearly exhibited a
faster than exponential growth as can be seen from the upward
curvature of the logarithm of the price as a function of time in
Figs. \ref{Figisr}-\ref{Fignica}, a fit of the price index time
series with expressions (\ref{eq:pure}) and (\ref{eq:tanh}) give
an exponent $\alpha$ larger than $15$ and critical times $t_c$ in
the range $2020-2080$, which are un-realistic. The results are not
improved by reducing the time intervals over which the fits are
performed. The results are not improved either by using the
alternative formula (\ref{eq:pure22}) valid for $m>2$ for which
the singularity is weaker as it occurs only on the slope of the
log-price.

It is possible that these problems stem from the fact that the
latter prices close to the end of the time series start to enter a
cross-over to a saturation, as would be expected due to
finite-size and rounding effects. Indeed, close enough to the
mathematically predicted singularity, one expects that the
realized price indexes will eventually saturate and the price
dynamics will enter another regime. We believe that it is the
start of this regime that makes difficult our recovery of the
parameters of expressions (\ref{eq:pure}) and (\ref{eq:tanh}). In
other words, the problem is not the difference between
(\ref{eq:pure}) and (\ref{eq:tanh}) capturing a non-critical
regime at early times but rather a cross-over to a saturation of
the singularity at the latest times.

Since we do not have a theory describing the saturation of the
super-exponential growth, we resort to the trick of fitting $p(t)$
rather than $\ln p(t)$ with the right-hand-sides of expressions
(\ref{eq:pure}) and (\ref{eq:tanh}). This procedure can be seen as
the continuous ODE formulation of the double-exponential
description of the price index growth advocated by Mizuno et al.
\cite{MTT02}. The results are shown in Figs. \ref{Figisr},
\ref{Figbra}, \ref{Fignica} and Table \ref{3badca}. Notice that,
in contrast with the previous cases of Bolivia, Peru, Hungary and
Germany, the characteristic cross-over time $\tau$ is rather
small, signaling the existence of a significant non-critical
regime at early times. For Israel, the fit of the price index
$p(t)$ with the right-hand-side of Eq.~(\ref{eq:tanh}) fails since
it has a much larger fit error $\chi=0.348$ than that $\chi=0.085$
of the fit using Eq.~(\ref{eq:pure}) and the estimated
$t_c=1991.44$ is too far off compared with $t_c=1988.06$ for the
fit of (\ref{eq:pure}).

\section{Conclusion}

We have presented a novel analysis extending the recent work of
Mizuno et al. \cite{MTT02} who analyzed the hyperinflations of
Germany (1920/1/1-1923/11/1), Hungary (1945/4/30-1946/7/15),
Brazil (1969-1994), Israel (1969-1985), Nicaragua (1969-1991),
Peru (1969-1990) and Bolivia (1969-1985). On the basis of a
generalization of Cagan's model of inflation based on the
mechanism of ``inflationary expectation'' or positive feedbacks
between realized growth rate and people's expected growth rate,
Mizuno et al. \cite{MTT02} have proposed to describe the
super-exponential hyperinflation by a double-exponential function.
Here, we have extended their reasoning by noting that the
double-exponential function is nothing but a discrete time-step
approximation of a more general nonlinear ODE formulation of the
price dynamics which exhibits a finite-time singular behavior. In
this framework, the double-exponential description is
undistinguishable from a power law singularity, except close to
the critical time $t_c$. Our new extension of Cagan's model, which
makes natural the appearance of a critical time $t_c$, has the
advantage of providing a well-defined end of the clearly
unsustainable hyperinflation regime. We have calibrated our theory
to the seven price index time series mentioned above and find an
excellent and reliable agreement for Germany (1920/1/1-1923/11/1),
Hungary (1945/4/30-1946/7/15), Peru (1969-1990) and Bolivia
(1969-1985). For Brazil (1969-1994), Israel (1969-1985) and
Nicaragua (1969-1991), we think that the super-exponential growth
is already contaminated significantly by the existence of a
cross-over to a stationary regime and the calibration of our
theory to these data sets has been more problematic. Nevertheless,
by a simple change of variable from $\ln p(t)$ to $p(t)$, we
obtain reasonable fits, but with much less predictive power. The
evidence brought here of well-defined power law singularities
reinforces the concept that positive nonlinear feedback processes
are important mechanisms to understand financial processes, as
advocated elsewhere for financial crashes (see
Ref.~\cite{bookcrash} and references therein) and for population
dynamics \cite{IS02,SI02}.

\textbf{Acknowledgments}

We acknowledge stimulating discussions with Mr. Takayuki Mizuno.
This work was partially supported by the James S. Mc Donnell
Foundation 21st century scientist award/studying complex system
(WXZ and DS).

\clearpage

\begin{table}
\begin{center}
\caption{\label{4function} Parameters of the fits with formulas
(\ref{eq:pure}) and (\ref{eq:tanh}) as indicated in the first
column to the hyperinflation price time series of the four
countries Bolivia, Peru, Hungary and Germany. The price indices
entered a crossover regime after the end of each period except for
Hungary whose hyperinflation of the Pengo was stopped artificially
by the introduction of the Forint in July 1946. $\chi$ denotes the
root-mean-square residue of the mean-square fits. The price index
$p(t)$ for Bolivia and Peru is normalized to $p(1969)=1$ at the
beginning of 1969. For Hungary, the time series is the price index. 
For Germany, it is the exchange rate between the Mark and the US dollar.}
\medskip
\begin{tabular}{llllllll}
\hline\hline Country&Period&$t_c$&$\alpha$&$\tau$&$A$&$B$&$\chi$\\\hline
Bolivia (\ref{eq:pure})&1969-1985&1986.94& 1.3&/ &-0.48&29.0&0.204\\
Bolivia (\ref{eq:tanh}) &1969-1985&1986.94& 1.3&1000&-0.49&0.068&0.205\\
Peru (\ref{eq:pure})&1969-1990&1991.29& 0.3&/ &-14.17&34.0&0.291\\
Peru (\ref{eq:tanh}) &1969-1990&1990.70& 0.01&22.03&-12050&12059&0.283\\
Hungary (\ref{eq:pure})&1945/04/30-46/07/15&46/09/03& 1.0&/ &-1.02&2370&1.168\\
Hungary (\ref{eq:tanh}) &1945/04/30-46/07/15&46/09/03& 1.0&1000&-1.68&2.69&1.177\\
Germany (\ref{eq:pure})&1920/01/01-23/11/01&23/12/18& 0.6&/ &-5.09&272&0.490\\
Germany (\ref{eq:tanh}) &1920/01/01-23/11/01&23/12/01& 0.3&945&-15.8&14.5&0.459\\
\hline\hline
\end{tabular}
\end{center}
\end{table}

\begin{table}
\begin{center}
\caption{\label{3badca} Parameters of the fits of the price index
$p(t)$ (and not of $\ln p(t)$) by the right-hand-side of formulas
(\ref{eq:pure}) and (\ref{eq:tanh}) as indicated in the first
column to the hyperinflation price time series of the three
countries Israel, Brazil and Nicaragua. The price indices entered
a crossover regime after the end of each period. $\chi$ denotes
the root-mean-square residue of the mean-square fits. The price
index $p(t)$ of each country is normalized to $p(1969)=1$ at the
beginning of 1969. }
\medskip
\begin{tabular}{llllllll}
\hline\hline Country&Period&$t_c$&$\alpha$&$\tau$&$A$&$B$&$\chi$\\\hline
Israel (\ref{eq:pure})&1969-1985&1988.06& 5.7&/ & 0.78&5.10E6&0.085\\
Israel (\ref{eq:tanh})&1969-1985&1991.44& 2.3& 2.20&-2.84E5&2.84E5&0.348\\
Brazil (\ref{eq:pure})&1969-1994&1997.50&16.3&/ & 1.93&3.66E21&0.604\\
Brazil (\ref{eq:tanh})&1969-1994&1994.97&18.6& 1.14&-5.77E9&5.77E9&1.196\\
Nicaragua (\ref{eq:pure})&1969-1991&1992.91&14.9&/ & 3.24&4.91E15&0.848\\
Nicaragua (\ref{eq:tanh})&1969-1991&1991.46& 9.2& 0.69&-8.03E8&8.03E8&0.945\\
\hline\hline
\end{tabular}
\end{center}
\end{table}

\clearpage

\end{document}